\documentclass[superscriptaddress,aps,pra,nolongbibliography,showpacs,twocolumn,floatfix]{revtex4-1} 

\usepackage{graphicx,helvet}
\usepackage{color}
\usepackage{bm}                      
\usepackage{tikz}
\usepackage{amsfonts}
\definecolor{mygray}{gray}{0.4}
\definecolor{light-blue}{rgb}{0.8,0.85,1}
\graphicspath{{../}}
\usepackage{amsmath}%
\usepackage{MnSymbol}%
\usepackage{wasysym}%
 \usepackage{soul}
\usetikzlibrary{decorations.shapes}

\newcommand{\beqa}{\begin{eqnarray}}
\newcommand{\eeqa}{\end{eqnarray}}
\newcommand{\beq}{\begin{equation}}
\newcommand{\eeq}{\end{equation}}

\usepackage{amsmath}
\DeclareMathOperator{\sign}{sign}

\def\inner#1#2{{\langle#1|#2\rangle}}

\newcommand{\ipr}{\mathrm{IPR}}

\newcommand{\tr}{\mathop{\mathrm{tr}}\nolimits}

\newcommand{\mean}[1]{\left\langle #1 \right\rangle}
\newcommand{\fmic}{f_{\Omega}(t)}
\newcommand{\fDR}{f_{\rm DR}}

\newcommand{\AFA}{{fidelity }}	
\newcommand{\ket}[1]{{\vert #1 \rangle}}
\newcommand{\bra}[1]{{\langle #1 \vert}}

\newcommand{\ifimar}{Instituto de Investigaciones F\'isicas de Mar del Plata (IFIMAR), CONICET-UNMdP,  Mar del Plata,  Argentina}
\newcommand{\conicet}{Consejo Nacional de Investigaciones Cient\'ificas y Tecnol\'ogicas (CONICET), Argentina}
\newcommand{\uba}{Departamento de F\'isica ``J. J. Giambiagi'' and IFIBA, FCEyN, Universidad de Buenos Aires, 1428 Buenos Aires, Argentina}
\begin{document}
\title{Lyapunov decay in quantum irreversibility}
\author{Ignacio Garc\'ia-Mata} \affiliation{\ifimar} \affiliation{\conicet}
\author{Augusto J. Roncaglia}\affiliation{\uba}
\author{Diego Wisniacki} \affiliation{\uba}

\begin{abstract}
The Loschmidt echo  -- also known as fidelity --
is a very useful tool to study irreversibility in quantum mechanics due to 
perturbations or imperfections.  Many different regimes, as a function of time and  strength 
of the perturbation, have been identified.
For chaotic systems, there is a range of perturbation strengths where the decay of the Loschmidt echo is 
 perturbation independent,  and given by the
classical Lyapunov exponent. 
But observation of the Lyapunov decay depends strongly on the  type of initial state upon which an average is done.
This dependence can be removed by averaging  the fidelity
over the Haar measure, and the Lyapunov regime is 
recovered, as it was shown for quantum maps. 
In this work we introduce an analogous quantity for systems with infinite dimensional Hilbert space, in particular 
the quantum stadium billiard, and we show clearly the universality of the Lyapunov regime.
\\ 

\noindent
{\bf Keywords: } Quantum Chaos, Irreversibility, Foundations of Quantum Mechanics
\end{abstract}
\pacs{03.65.Yz, 03.65.Ta, 05.45.Mt}
%
%
\maketitle
\section{Introduction}
Understanding the emergence of irreversibility from the basic laws of physics has 
been a longstanding  problem.  Although, since %
the eighteenth century 
it is known that the second law describes the arrow of time, its microscopic foundation 
has been matter of debate until these days
\cite{Fermi,Maxweedemon}. The main problem is that classical mechanics is
time-symmetric and cannot explain the emergence of the thermodynamic arrow of time.
This contradiction has been apparently resolved with the understanding of chaos. The sensitivity to 
initial conditions of chaotic systems, along with the notions of mixing and coarse graining, has been the main 
argument to explain irreversibility in classical systems \cite{Gaspard-book}.

In quantum mechanics, the situation is more involved.
Due to the linearity of the Schr\"{o}dinger  equation
there is no sensitivity to initial conditions, and  
therefore the origin of irreversibility in quantum mechanics lies elsewhere.
For this reason, 
an alternative idea was proposed by Peres \cite{Peres1984}. 
He suggested that quantum mechanics is sensitive to perturbations in the evolution rather than
to the initial conditions.
A suitable dynamical quantity to study such a behaviour was coined fidelity or Loschmidt echo (LE), which is
defined as,
\beq
M_{\psi}(t)=|\bra{\Psi} U^\dagger_{\xi+\delta\xi} (t) U_{\xi} (t)\ket{\Psi}|^2
\label{LE}
\eeq
where $U_\xi(t)$ is an evolution operator and,   $U_{\xi+\delta\xi}(t)$
is a corresponding perturbed one,  $\ket \psi$ an initial state, and the parameter $\delta\xi$ characterises the 
strength of the perturbation. Thus, Eq.~(\ref{LE})
can be interpreted in two different ways. 
On the one hand,
it is the overlap between an initial state $\ket \psi$ evolved
forward up to time $t$ with the evolution operator $U_\xi(t)$, and the same state evolved backward in
time with a perturbed evolution operator $U_{\xi+\delta\xi}(t)$. 
On the other hand,  
it can also be interpreted as the overlap at time
$t$ of the same state evolved forward in time with slightly different Hamiltonians. 
While the first interpretation
gives the idea of irreversibility, the second is related to the 
sensitivity to perturbations of quantum evolutions.

The LE has been intensively studied in the first decade of this century \cite{Gorin2006,Jacquod2009,DiegoScholar}, and
several time and perturbation regimes were clearly identified using different techniques like random matrix theory, semiclassical, and numerical simulations \cite{DiegoScholar}.  
 The progress in experimental techniques has permitted to study the LE in various different settings like NMR \cite{PastawskiRMN}, microwave billiards\cite{expbilliards}, elastic waves \cite{lobkis} and cold 
atoms \cite{coldatoms}.
The most relevant result in connection with chaos and irreversibility is that 
the LE has a regime where the decay becomes independent of the
perturbation strength and it is given by the Lyapunov exponent \cite{Jalabert2001}, 
a classical measure of the divergence of neighbouring trajectories \cite{Gaspard-book}.  
The Lyapunov regime has been observed in several systems \cite{Jacquod2001,PhysRevE.65.055206, PhysRevE.68.056208,Cucc2004,PhysRevE.70.026206, PhysRevE.67.056217,Wang2004, Pozzo2007}. 
In these works a crucial feature is that the initial states need to be coherent (Gaussian) wavefunctions \cite{Diego2002}.  
Besides, an average on initial condition or perturbations is required.

The dependence of the LE with the type of initial state can be removed by considering
an average over initial states according to the Haar measure for finite
dimensional systems \cite{zanardi2004} %
\begin{equation}
\label{Haar}
\int d\ket{\Psi} \, M_{\psi}(t) = \frac{1}{d(d+1)}[d+|\mean{U_\xi(t), U_{\xi+\delta\xi}(t)}_{HS}|^2] ,
\end{equation}
where  $\mean{U_\xi(t), U_{\xi+\delta\xi}(t)}_{HS}\equiv\tr[U^\dagger_\xi(t) U_{\xi+\delta\xi}(t)]$ is the Hilbert-Schmidt product between the operators,
and $d$ is the dimension of the Hilbert space. Thus,
the   average fidelity amplitude
\begin{equation}
|f(t)|=|\tr[U^\dagger_\xi(t) U_{\xi+\delta\xi}(t)]|
\label{eq:fid}
\end{equation}
 which is directly related to the LE, is a  state-independent quantity. 
This quantity was studied in detail for quantum maps \cite{garciamaNJP}.
Analytical results were obtained using a semiclassical theory known as dephasing representation (DR) \cite{vanicek2003,vanicek2004,vanicek2006}.
It is shown that $ |f(t)|$ has two clear decay regimes. For short times, the decay rate depends on the 
perturbation and it is predicted considering random dynamics. This corresponds to the limit of infinite Lyapunov exponent. If the strength of the perturbation is small enough, this regime lasts up to the saturation point.  The other regime,  was obtained considering  that the perturbation is completely random. That is, after each step of the map, the perturbation contributes with a random phase to each trajectory. In that case, using the DR and transfer matrix theory it is shown that the asymptotic decay rate of  $|f(t)|$ 
is controlled by the largest classical Lyapunov exponent $\lambda$. 
Numerical tests of the analytical predictions were given for the quantum baker and a family of perturbed cat maps (see Sect.~\ref{quantum maps} below).

In this work we go one step further by studying $ |f(t)|$ in a realistic system. We consider a particle inside a stadium billiard that is perturbed
by a smooth potential consisting of a number of Gaussians randomly distributed inside the cavity.  
A 2-D billiard has an infinite-dimensional
Hilbert space. For this reason, instead of computing Eq.~(\ref{eq:fid}) for a
complete set,
we consider an initial state defined as an incoherent sum of all the energy projectors from the ground state up to a 
given high energy level. 
We have numerically computed this quantity that we call  $|\fmic|$ for the quantum stadium billiard
and show that 
it has similar behaviour to that observed  for quantum maps.
For short times, $|\fmic|$ has a decay that depends on the perturbation strength. 
But, after a crossover and for sufficiently large perturbation strength
we can clearly see that $|\fmic|$ decays exponentially with a decay rate given by the Lyapunov exponent of the classical billiard.
 In order to confirm these results, we have also computed $|\fmic|$ using the DR.
 We also show that the DR describes very well
the quantum behaviour and that the Lyapunov regime is also clearly observed in this approximation.

The rest of the paper is organised as follows. In Sec. \ref{quantum maps} we summarise the results obtained in Ref. \cite{garciamaNJP}
in quantum maps. We show that the DR works very nicely to describe the quantum behaviour and we show the different decay regimes
of  $ |f(t)|$. In  Sec. \ref{stadium} we show the $ |\fmic|$ for the stadium billiard. 
Final remarks and outlook are given in Sec. \ref{finalremarks}.
\section{quantum maps}
\label{quantum maps}
For this work to be self contained, in this section we briefly review the results previously 
obtained in \cite{garciamaNJP}  for quantum maps 
on a two dimensional phase space with periodic boundary conditions (2-torus). 
These maps have the essential ingredients of chaotic systems and are simple to treat numerically and sometimes even analytically.
The torus geometry implies that, upon quantisation, position $q$ and momentum $p$ are discrete and related by the discrete Fourier transform (DFT). The Hilbert space then has finite dimension $N$, and the semiclassical limit is given by $N\to \infty$. An efficient Planck constant can be thus defined as $\hbar=1/(2 \pi N)$.

As a tool we use the DR \cite{vanicek2003,vanicek2004,vanicek2006} which
avoids some of the drawbacks of other semiclassical methods.
The fidelity in the DR can be written as 
\begin{equation}
	\label{eq:odr}
\fDR(t)=\int dq dp W_{\rho}(q,p) \exp(-i \Delta S_{\delta\xi}(q,p,t)/\hbar),
\end{equation}
where $W(q,p)$ is the Wigner function of the initial state $\rho$, and
\begin{equation}
\Delta S_{\delta\xi}(q,p,t)=-\delta\xi\int_0^t d\tau V(q(\tau),p(\tau)) 
\end{equation}
is the action difference evaluated along the unperturbed classical trajectory. If $\rho$ is a maximally mixed state
then  Eq.~\ref{eq:odr} is an average over a complete set (and becomes basis independent),
\begin{equation}
	\label{eq:afa}
\fDR(t)=
\frac{1}{\cal V}\int dq dp \exp(-i \Delta S_{\delta\xi}(q,p,t)/\hbar),
\end{equation}
this quantity is the semiclassical  expression for the fidelity given in Eq.~(\ref{eq:fid}). 
Here for simplicity we write phase space variables $q$ and $p$, and their differentials, as one dimensional. But Eq.~(\ref{eq:afa}) holds for arbitrary dimensions. 
For maps time is discrete so, for the reminder of this section, we define $t:=n$, with $n$ integer. 

We use the DR to study the decay of the \AFA in the chaotic regime as follows. First we suppose that 
the system is very strongly chaotic, $\lambda\to \infty$. This is essentially equivalent to assuming that the evolution is
random without any correlation. In order to compute $f(n)$ we partition the phase space into $N_c$ cells and consider that the probability of jumping from one cell to the other is uniform. It can then be shown that
\begin{eqnarray}
\fDR(n)&=&\frac{1}{N^n}\sum_{j_1}\ldots\sum_{j_n}\exp[-i (\Delta S_{\delta\xi,j_1}+\ldots+\Delta S_{\delta\xi,j_n})/\hbar]\nonumber\\
&=&\left[\frac{1}{N^n}\sum_j \exp(-i\Delta S_{\delta\xi,j}/\hbar)\right]^n,
\end{eqnarray}
where $\Delta S_{\delta\xi,j_k}$ is the action difference on the cell $j$ at time $k$. Taking the limit $N_c\to\infty$, we get
\begin{equation}
\fDR(n)=\left[\int dq dp\exp(-i \Delta S_{\delta\xi}(q,p))\hbar\right]^n.
\end{equation}
The absolute value of $\fDR(n)$ can then be written as %
\begin{equation}
|\fDR(n)|=\exp(-\Gamma\, n),
\end{equation}
where
\begin{equation}
\label{eq:gamma2}
\Gamma=-\log \Big|\int \exp [ -i  \Delta S_{\delta\xi}(q,p)/\hbar] dq dp\Big| \, .
\end{equation}
Then, if the dynamics is completely random, which is approximately the case for strongly chaotic systems, then the \AFA
decays exponentially with a rate $\Gamma$.  As we shall see, this decay also explains the short time behavior regardless of $\lambda$ because for short times the dynamics can always be supposed to be uncorrelated.

To unveil the intermediate time regime we consider the limit of random perturbation. In \cite{garciamaNJP}, using the DR it is shown that, for the baker map with a random perturbation, the \AFA can be written as a sum of products of transfer matrices
\begin{equation}
\label{fidmat}
\fDR(n)=\frac{1}{2^{n/2+L-1}}\sum_{k0,\ldots,k_n}M_{k_0,k_1}\ldots M_{k_{n-1},k_n},
\end{equation}
where
\begin{equation}
k_i=2^{(L-1)}\times \cdot\mu_i\ldots\mu_{L+i-2},
\end{equation}
the digits $\mu_i=0,1$ define position and momentum
 \begin{eqnarray}
 q&=&\sum_{j=0}^{\infty} \frac{\mu_j}{2^j+1}\stackrel{\rm def}{=}\cdot\mu_0\mu_1\ldots\\
 p&=&\sum_{j=0}^{\infty} \frac{\mu_{-j}}{2^j+1}\stackrel{\rm def}{=} \cdot\mu_{-1}\mu_{-2}\ldots\\
 \end{eqnarray}
 in symbolic dynamics (see e.g. \cite{Saraceno1994}). A point in phase space is then
 $(q,p) = \ldots \mu_{-2}\mu_{-1}\cdot\mu_{0}\mu_{1}\ldots$, and one step of the map consists in shifting the point to the right. The letter $L$ in the previous equations indicates a truncation size of the symbolic dynamics expansion.
 Defining the unit norm vector $\ket{1}=2^{-(L-1)/2}(1,1,\ldots,1)$,  
 Eq.~(\ref{fidmat}) can be written in compact form as
 \begin{equation}
 \fDR(n)=2^{-n/2}\bra{1}M^n\ket{1}
 \end{equation}
The properties of the \AFA are then determined by the spectrum of the finite matrix $M$. In particular the asymptotic decay is ruled by the largest eigenvalue (in modulus) of $M$. Considering the special structure of the transfer matrices for the bakers map, it was shown that %
\begin{equation}
|\fDR(n)|^2\approx 2^n=e^{\lambda_B n}
\end{equation}
where $\lambda_B=\ln 2$ is the largest Lyapunov exponent of the baker map. This analytical result was further extended to more general types of maps \cite{garciamaNJP}.

We show numerically these two regimes for a family of perturbed cat maps  \cite{dematos}
\begin{equation}
\left.
\begin{array}{ccl}
\label{eq:pcat}
\bar{p}&=&p-a\,q+\xi f(q)\\
\bar{q}&=&q-b \bar{p}+\tilde{\xi} h(\bar{p})
\end{array}
\right\}\quad ({\rm mod}\  1).
\end{equation}
For simplicity let $\tilde{\xi}=\xi$.
For $a,b$ integer, these maps are uniformly hyperbolic and for small enough  $K$  
the Lyapunov exponent is approximately given by
 \begin{equation}
\lambda \approx \log \frac{1 + ab + \sqrt{ab(4+ab)}}{2} \, .
\end{equation}
\begin{figure}[h]
\includegraphics[width=0.95\linewidth]{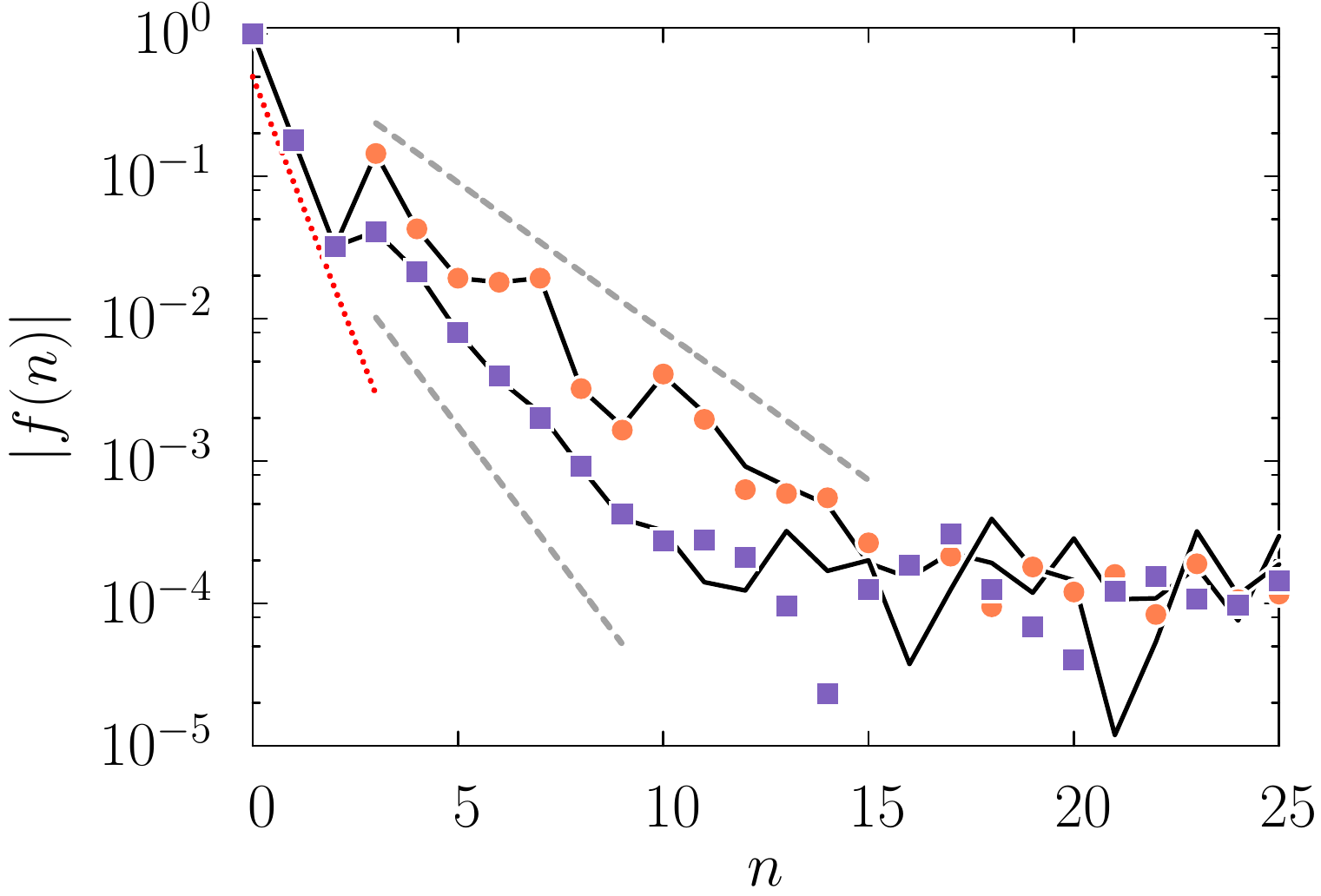} 
\caption{Fidelity as a function of discrete time $n$ 
using straight quantum calculation (points) and DR (black solid line)
for the perturbed cat map, with $a=1$ (circles), corresponding to $\lambda\approx 0.962$ and 
$a=2$ (squares) corresponding to $\lambda\approx 1.76$. Here $N=10^4 $ and $\xi=0.025$ and $\delta \xi/\hbar=2$. The dashed (gray) lines correspond to the Lyapunov regime, $|f(n)|\sim \exp(-\lambda n/2)$. The dotted (red) line marks the short time decay, \mbox{$|f(n)|\sim \exp(-\Gamma n)$}, with  $\Gamma$ obtained from 
Eq.~(\ref{eq:gamma2}).
\label{fig:mapQVani}}
\end{figure}
For simplicity from now on we take $a=b$.
These maps can be written in the general form 
\begin{equation}
\left.
\begin{array}{ccl}
\bar{p}&=&p-\frac{dV_{\xi}(q)}{dq}\\
\bar{q}&=&q +\frac{dT_{\xi}(\bar{p})}{d\bar{p}}
\end{array}
\right\} \quad ({\rm mod}\  1).
\end{equation}
and can be simply quantized as a product of two operators
\begin{equation}
\label{kicked}
U_\xi=e^{-i 2\pi N T_\xi (\hat{p})}{\rm e}^{-i 2\pi N V_\xi(\hat{q})}.
\end{equation}
Many well known quantisations of classical maps can be expressed in this way, e.g. the kicked Harper map \cite{Leboeuf1990}, and the Chirikov standard map \cite{DimaScholar}.
For the numerical examples 
we consider
\begin{eqnarray}
\label{eq:pert}
f(q)&=&2 \pi \left[ \cos(2 \pi q) - \cos(4 \pi q) \right] \, , \\ 
h(\bar{p}) &=&0 \, .
\end{eqnarray} 
 as the perturbing ``forces'' of Eq.~(\ref{eq:pcat})

In Fig.~\ref{fig:mapQVani} we show two things. On the one hand, the almost prefect agreement of the DR calculation of $|f(t)|$ against the straightforward quantum result. On the other hand, it is shown that the two different exponential regimes can be distinguished. For the sake of clarity we show results for $a=1$ and $2$ which correspond to $\lambda\approx 0.962$ and $1.72$ respectively. 

\begin{figure}
\includegraphics[width=0.95\linewidth]{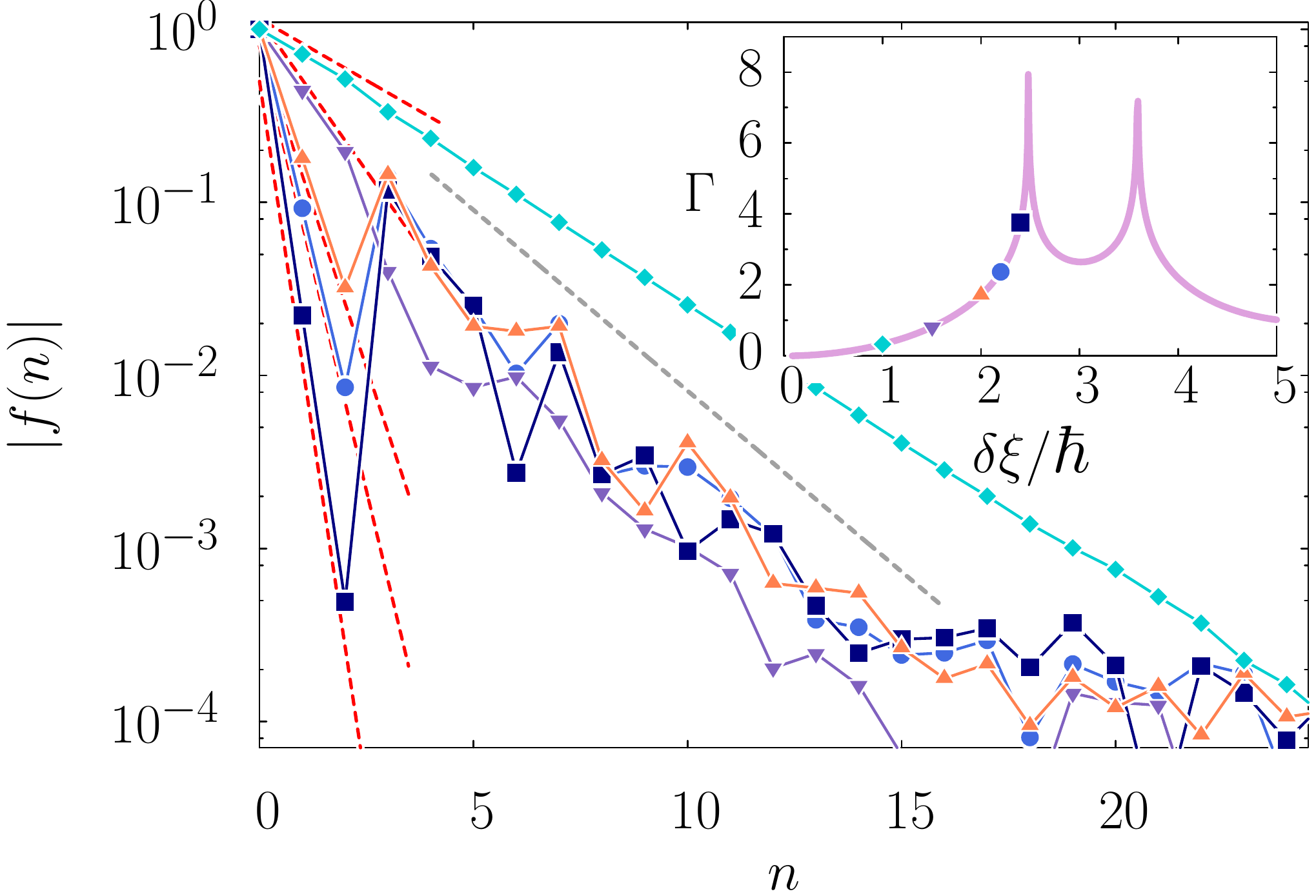} 
\caption{\label{fig:mapObj}
Quantum calculation of the \AFA  for the perturbed cat map with $a=1$ and 
different values of the perturbation, $\delta\xi/\hbar=1 (\diamond),\,1.5 (\triangledown),\, 2 (\triangle),\, 2.2 (\circ),\, 2.4 (\square)$. 
The dashed red lines show the small time behavior $|f(t)|\sim \exp(-\Gamma n)$. The values
of $\Gamma$ obtained from Eq.~(\ref{eq:gamma2}) are shown in the inset.}
\end{figure}
In Fig.~\ref{fig:mapObj} we show a detailed example illustrating how different the two regimes can be. There, five examples of 
$|f(t)|$ for different values of $\delta K$ are displayed. It can be clearly observed that the initial decay rate
is given  by $\Gamma$. In the inset we show $\Gamma$ as a function of the perturbation and the points mark the decay rate value indicated by the dashed (red lines) in the main panel. After this short time decay there is a revival and then the fidelity again decays exponentially with a rate given by $\lambda$, except in the case where $\Gamma \ll \lambda$.

From the evidence of Figs.~\ref{fig:mapQVani} and \ref{fig:mapObj} a behaviour like
\begin{equation}
|f(t)| \sim \exp(-\Gamma t)+A \exp(-\lambda t/2) 
\end{equation}
can be hinted. The decay given by the rate $\Gamma$ is explained by an initial lack of correlations. If the dynamics is strongly chaotic, then this is the decay that dominates throughout the evolution. This can be simulated by random evolution.
In other cases there is a crossover from to the perturbation independent Lyapunov regime. 
In \cite{garciamaNJP} a random perturbation model was used to demonstrate this crossover, and also the crossover time could be inferred.

\section{Stadium billiard}
\label{stadium}
In the previous sections we show that  $|f(t)|$ is a suitable quantity to characterise quantum irreversibility in d-dimensional systems. 
It does not depend on the initial
conditions because it is the trace of the echo operator $U^\dagger_{\xi+\delta\xi} (t) U_{\xi} (t)$.  
Moreover, in the case of abstract maps,
the DR can be used to show analytically that there is a Lyapunov regime that
does not depend on the type of initial states, contrary to what happens in the case of the LE.      
Now we will study the behaviour of a similar quantity in a realistic system, a particle inside a billiard. In this system, the Hilbert space is infinite-dimensional and it is not
possible to compute the trace of the echo operator. For this reason, we consider
\beq
|\fmic|=|tr[U^\dagger_{\xi+\delta\xi} (t) U_{\xi} (t) \rho_{\Omega}]|
\label{fiomega}
\eeq
where the initial density function $\rho_{\Omega}(m)=m^{-1}\sum_{i=0}^{m} |E_i(\xi) \rangle \langle  E_i(\xi) |$, 
a microcanonical state located in an energy window
that start in the ground state up to the  $m$-th excited state. 
We note that $|\fmic|$ is related to a well known quantity in non-equilibrium statistical mechanics: the 
probability of doing work $W$.
This can be seen by considering a system with Hamiltonian 
$H(\xi)$ that is in an initial equilibrium state $\rho$. At $t=0$ the energy is measured and a quench $H(\xi) \rightarrow H(\xi+\delta\xi)$ is done.
Then,  the system evolves a time $t$ and another energy measurement is done.  If $E_i(\xi)$ and $E_j(\xi+\delta\xi)$ are the results of the measurements, the work done on the system is  
$W=E_j(\xi+\delta\xi)-E_i(\xi)$.  Then, it is easy to show that the probability of work $P(W)$ is the Fourier transform of
\beq
f_{\beta}(t)=tr[U^\dagger_{\xi+\delta\xi} (t) U_{\xi} (t) \rho],
\eeq
here we consider the absolute value of this quantity Eq. \ref{fiomega}.

\begin{figure}
\includegraphics[width=0.95\linewidth]{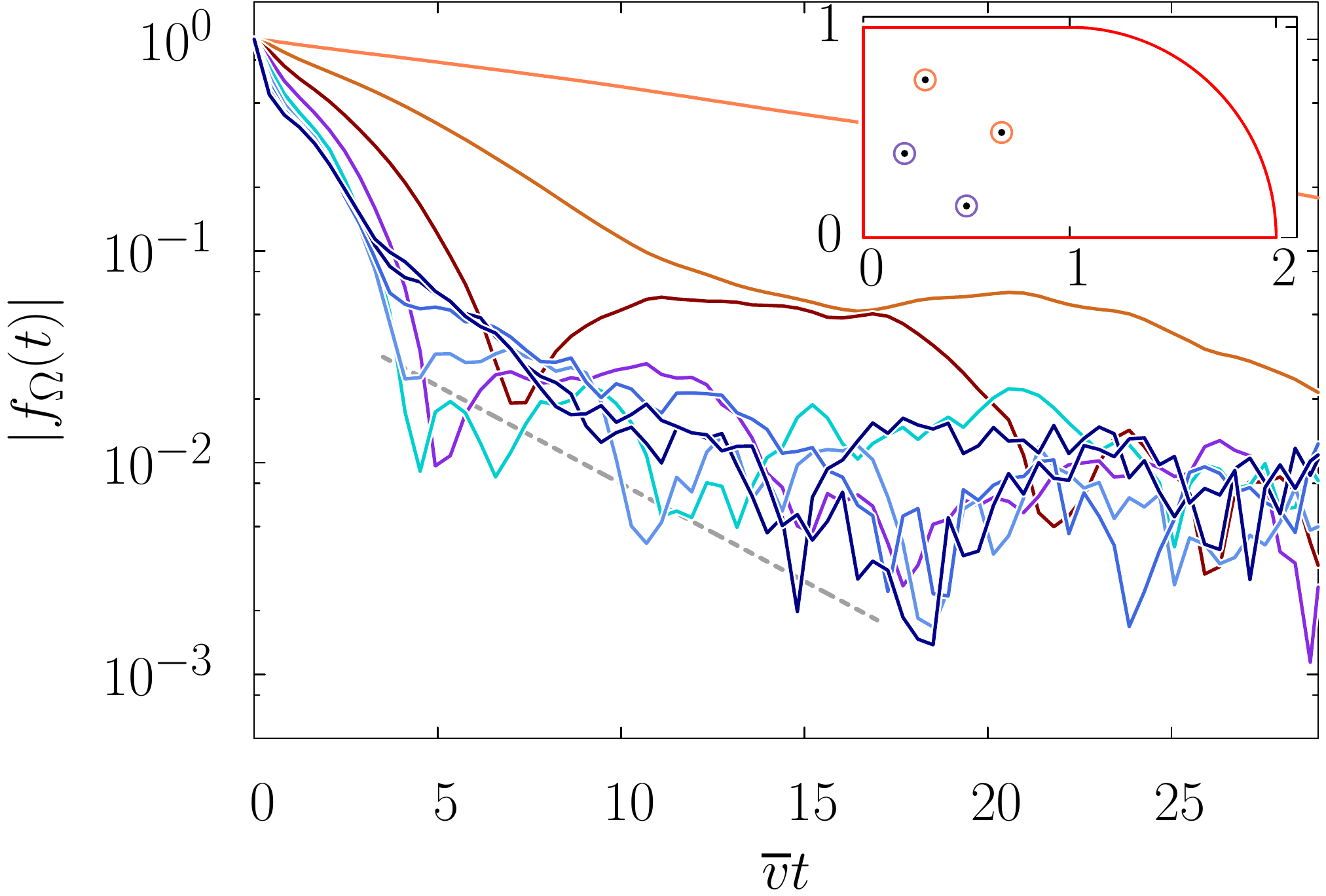} 
\caption{Fidelity for the stadium billiard with initial state $\rho_\Omega({m=500})$ and a  basis of $3134$ states, as a function of the rescaled time $\overline{v} t$.   
$\delta\xi=10,\, 30,\, 50,\, 70,\, 80,\, 90$. In the inset we show the position of the four Gaussians and 
diameter of the circles is $\sigma=0.1$ (blue indicate positive Gaussian and red negative).
The dashed grey line is $ \exp(-\lambda_1t)$. \label{afaStad}}
\end{figure}
We have studied $|\fmic|$ for a particle 
in the desymmetrised stadium billiard with radius $r=1$ and straight line of length $l=1$ (see inset of Fig. \ref{afaStad}). 
The perturbation  is a smooth potential consisting in a series of four Gaussians, 
\beq
V(x,y,\delta \xi)=\delta \xi \sum_{i=1}^4 \sign_i \exp(-[(x-x_i)^2-(y-y_i)^2]/(2 \sigma^2)]
\label{pot}
\eeq
with $\delta \xi$ the perturbation strength, $\sigma=0.1$ their widths,  
$(x_1,y_1)=(0.2,0.4)$, $(x_2,y_2)=(0.67,0.5)$, $(x_3,y_3)=(0.5,0.15)$ and  $(x_4,y_4)=(0.3,0.75)$  the positions of the centres and $\sign_1=\sign_3=1$, $\sign_2=\sign_4=-1$. The eigenstates of the unperturbed stadium ($\xi=0$) are obtained 
using the scaling method \cite{vergini95}. The states of the perturbed system are obtained by diagonalising the perturbed Hamiltonian in the basis of the stadium billiard. We have used several number of unperturbed states to check the convergence of the results.  We point out that the perturbations considered in this work and in \cite{garciamaNJP} affect the whole phase space. There has been  some theoretical \cite{Goussev2008,Ares2009,Hohmann,Kober} and experimental  
efforts to study the effect of local perturbations.  In theses works they either  find  a crossover from a Fermi-golden-rule regime to an exponential regime with rate given by the so-called escape rate (given by a representative size of the perturbation) \cite{Goussev2008,Kober}, or an algebraic decay regime \cite{Hohmann}.
But, even though they also consider the average fidelity amplitude, they do not find a Lyapunov regime.Therefore in this work we only consider global perturbations.     
\begin{figure}
\includegraphics[width=0.95\linewidth]{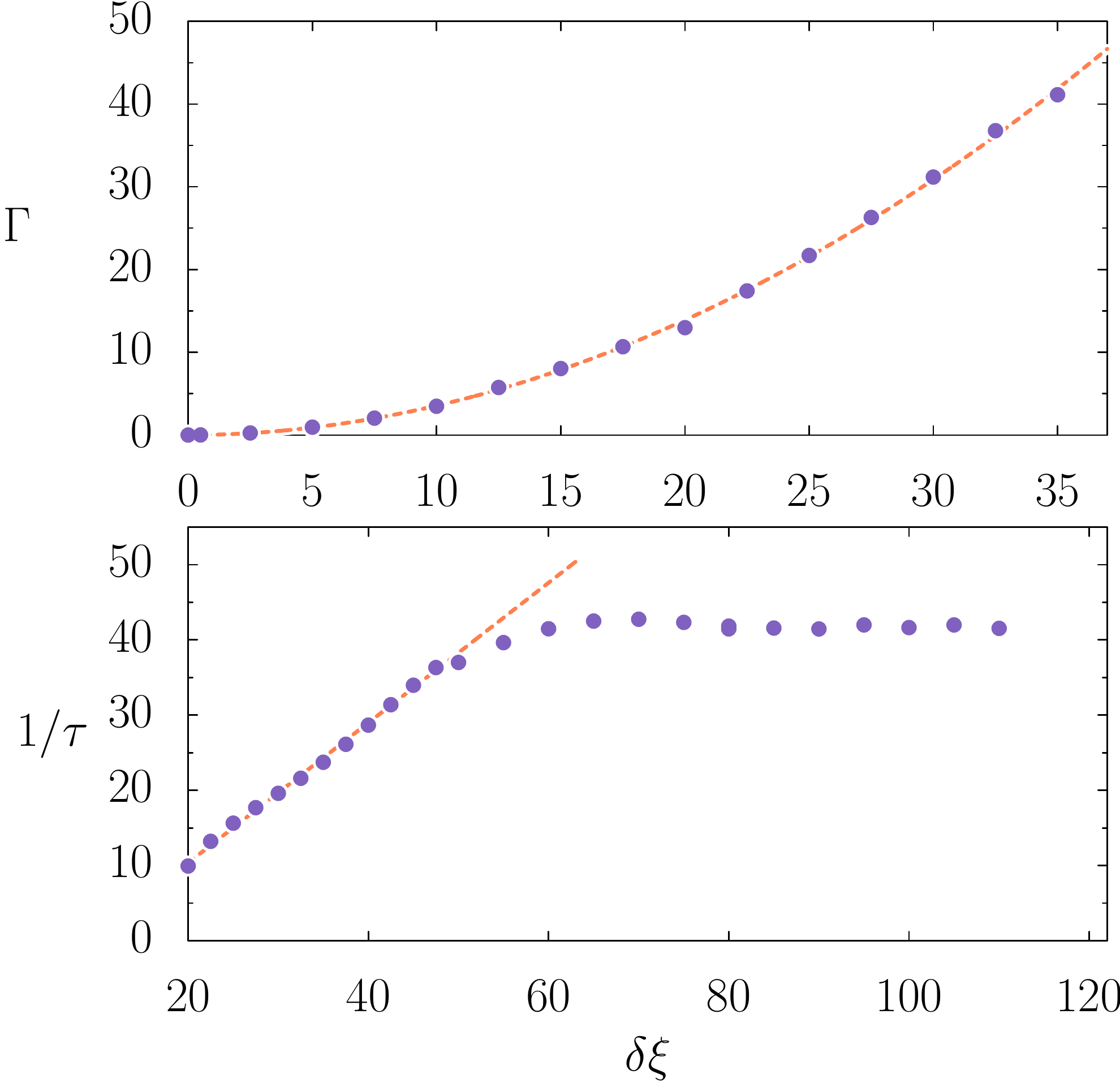} 
\caption{(top) Decay rate $\Gamma$ for the short time decay in the perturbative regime, $|\fmic|\sim \exp[-\Gamma \,t]$. As expected from the
Fremi-Golden rule $\Gamma$ has approximately quadratic dependence with $\delta\chi$ (the dashed red line
is a fit where $\Gamma\propto \delta\chi^{1.969\pm 0.034}$ \label{Short_t}.
(bottom) Characteristic time $1/\tau$ for the Gaussian decay $|\fmic|\sim \exp[-(t/\tau)^2]$, for small $t$. 
The slope of the linear fit is $0.93\pm 0.013$, so $1/\tau\approx \delta\chi$ until saturation.
Initial state is $\rho_\Omega({n=800})$  and the basis used has 3134 states, the width of the Gaussian perturbation is $\sigma=0.1$.}
\end{figure}

In Fig.~\ref{afaStad} we show $|\fmic|$ for the stadium billiard. The unperturbed evolution is given by the free dynamics inside the cavity ($\xi=0$) and the perturbed one is 
with the potential of Eq.~(\ref{pot}). Results for several perturbation strengths $\delta\xi$ are shown.  
The initial microcanonical state corresponds to the first 500 
eigenstates  of the unperturbed system. The calculations were done using the first 3135 
eigenstates of the stadium billiard. The convergence of the results were tested using
a bigger basis of up to 5600 states. For a smaller basis of 1300 states the results are also well behaved.

Let us first analyse the small time behaviour.
For small $\delta\xi\lesssim 25$, $|\fmic|$ decays exponentially $\sim \exp(\Gamma t)$. 
In Fig.~\ref{Short_t} (top) we show $\Gamma$ as a function of $\delta \xi$. 
As expected, we can clearly see that $\gamma \sim \delta \xi^2$. Such behaviour is referred to as the Fermi-golden-rule regime \cite{Jacquod2001}. 
For $\delta\xi\gtrsim 25$, the short time decay  $|\fmic|$ is approximately a Gaussian
function $\sim \exp[ -t^2/(\tau^2)]$ (see Fig.~\ref{afaStad}).  
In Fig.~\ref{Short_t} (bottom) we show 
$1/\tau$ as a function of $\delta \xi$. We can see that after a transient, there is a region where  
$1/\tau$ does not depend on the perturbation strength $\delta \xi$.

\begin{figure}
\includegraphics[width=0.95\linewidth]{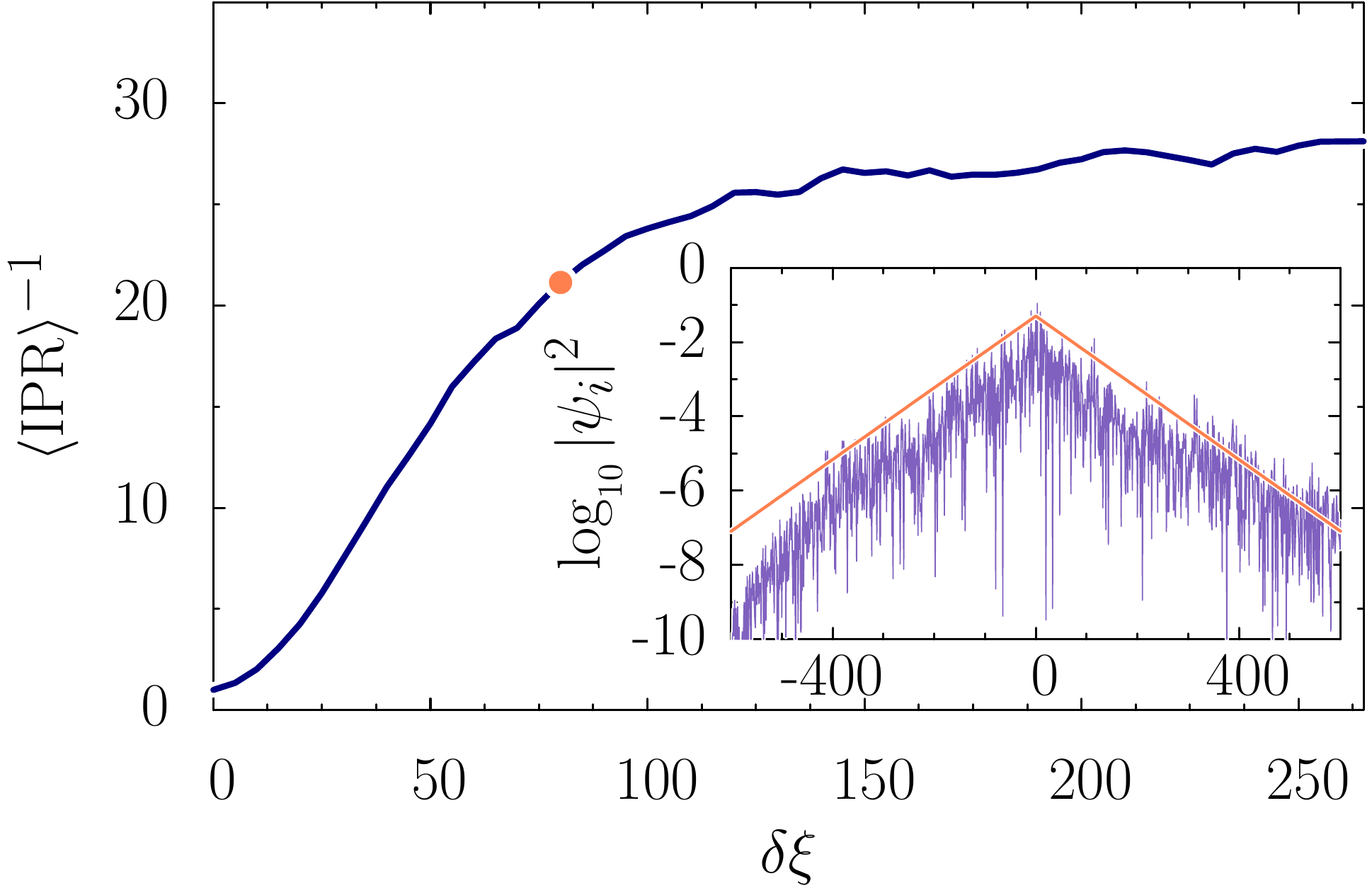} 
\caption{Inverse of the average IPR for the first 800 perturbed states as a function of $\delta \xi$. 
Inset:  $|\ket{E_j}_i|^2=|\inner{E_j(\xi+\delta\xi)}{E_i(\xi)}|^2$ for the state $j=600$ for $\delta \xi=80$. \label{ipr}}
\end{figure}

The behaviour of the fidelity $|\fmic|$ can be related to the spread of the initial state in the perturbed basis.
The natural quantity to study this type of localisation properties is the inverse participation ratio (IPR)  \cite{Fyodorov1995,Jacquod1995,Georgeot1997, Berkovits1998, Flambaum1994, Flambaum2001,PhysRevE.91.010902}.  
The IPR of a perturbed eigenstate $\ket{j(\xi+\delta\xi)}$ 
in the unperturbed basis $\ket{E_i(\xi)}$ is 
\begin{equation}
\ipr(\ket{E_i})=\left(\sum_m | \inner{E_m(\xi)}{E_i(\xi+\delta\xi)}|^4\right) 
\end{equation}
(throughout this section $\xi=0$).
The inverse of this quantity -- also called the participation number-- gives an estimation of the number of unperturbed states contributing to a given perturbed state. 
In Fig.~\ref{ipr} we show the inverse of the IPR  averaged over the first 800 states, as a function of $\delta \xi$.
We can see that it has an approximately quadratic growth up to $\delta \xi=20$. 
After that
the inverse of the  IPR grows linearly in the interval $20\lesssim \delta \xi\lesssim 60$. Finally the growth rate tends to a saturation at value which is much smaller than 
the basis size. This shows that the perturbed states remain localised in energy. 
As an example, in the inset of Fig.~\ref{ipr} we show the $|\psi_i|^2=\inner{i(\xi+\delta\xi)}{i(\xi)}|^2$ for the state corresponding to 
the level $600$ for $\delta \xi =80$ (marked by a red dot on the main panel).  The exponential decay 
of the tails is a manifestation of its localisation.  This localisation is responsible the plateau of $1/\tau$ shown in Fig.~\ref{Short_t} (bottom).
We remark that the three perturbation regimes of the short time decay of $\fmic $ [see Figs. ~\ref{afaStad} and ~\ref{Short_t}]
are manifested in the IPR behaviour [Fig. ~\ref{ipr}].

After the Gaussian decay shown in Fig.~\ref{afaStad}, we can see a second exponential decay with a rate given by the classical Lyapunov exponent of the stadium billiard.  
The Lyapunov exponent is  $\lambda=\lambda_1\overline{v}$, where $\lambda_1=0.43$ 
corresponds to $l=r=1$ \cite{Benettin1978,Dellago1995}. 
Here $\overline{v}=2\overline{k}$ is the average velocity computed from the eigenenergies $E_i=k_i^2$  ($k_i$ being the wavenumber
of the eigenstate $\ket{E_i}$) in the energy window $\Omega$ considered. 
Evidently, the fidelity computed using an initial state $\rho_\Omega$  and the fidelity obtained 
from the Haar measure for the quantum maps share the same decay behavior. 
A short time decay which depends on the characteristics and strength of the perturbation, 
followed by a Lyapunov regime depending on a classical feature.

\begin{figure}
\includegraphics[width=0.95\linewidth]{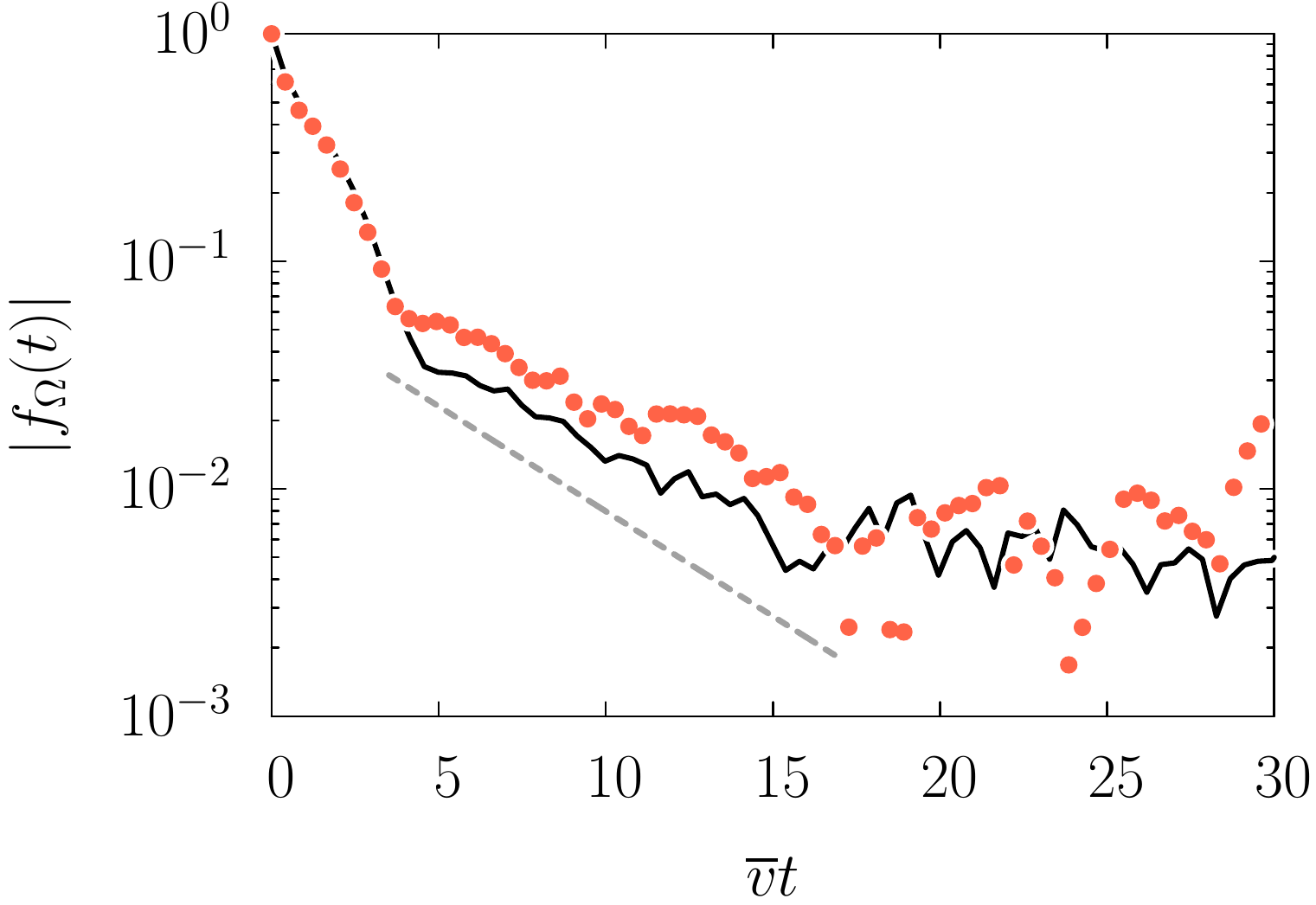} 
\caption{ $|f_\Omega (t)|$ for the stadium billiard with perturbation strength $\delta\xi=70$ and width $\sigma=0.1$. The (red) circles correspond to the quantum evolution, with initial state $\rho_\Omega (m=500)$ and  a basis of $3134$ states.
The solid line corresponds to DR calculation using 
$189991$ initial conditions with energies chosen 
 in the range corresponding to the first $500$ quantum energies.  The dashed gray line corresponds to the Lyapunov decay.
\label{StadVani}}
\end{figure}
We also compute $|\fmic|$ using the semiclassical DR.  This is a simple task due to the fact that unperturbed trajectories are geometrically obtained 
in the billiard and the perturbation only gives a phase as dictated by Eq. \ref{eq:odr}.
To take into account the  initial $\rho_\Omega(m)$ state, we compute the semiclassical $|\fmic|$ using that 
the initial conditions are uniformly random inside the billiard,
the same was assumed for the direction of the initial momentum. The modulus squared of the momenta  
are distributed as the eigenenergies of the unperturbed system. In Fig.~\ref{StadVani}
we show  $|\fmic|$  computed using the DR and the quantum results. We can see that the semiclassical approximation 
provides an accurate fitting of the quantum results. Moreover, the Lyapunov decay is clearly observed in the DR approximation.
\section{Conclusions}
 From the outset, the LE emerged as a viable quantity to characterise instability and irreversibility in quantum systems.
 A large amount of work was dedicated to describe the different regimes depending on the perturbation strength, but it received
 a real important boost  when the Lyapunov regime was first described linking classical an quantum chaotic behaviour.
 However although it was shown to exist in many different systems, all the semiclassical and numerical calculations showed 
 that the Lyapunov regime could only be observed if the initial states considered were ``classically meaningful'' 
 \cite{Jacquod2009} (typically coherent states).
 We found a solution to this problem considering the average fidelity amplitude, a basis independent quantity,
 which is closely related to the LE if one
 considers an average over the Haar measure (Eqs.~(\ref{Haar}) and (\ref{eq:fid})). 
Indeed, a recent work \cite{garciamaNJP},
  briefly reviewed in Sect.~\ref{quantum maps}, shows analytically and numerically that for quantum maps on the torus
 the average fidelity amplitude decays as a double exponential, where the first decay 
 rate is depends on the strength and type of perturbation, whereas the second decay rate is given by the classical Lyapunov regime.
 
But, although quantum maps have some generic properties of quantum chaos, they are not very generic systems themselves. So the challenge in this work was to take the analysis one step further and study the average fidelity amplitude in a more realistic system, paradigmatic of quantum chaos, like the stadium billiard. To overcome the problem of 
the infinite dimensional Hilbert space, where averaging over the Haar measure is unfeasible, 
we introduced an energy cutoff and considered the system 
to be initially in a state that has an equiprobable distribution over some energy window, in the same spirit of a microcanonical ensemble. 
In this way were able to recover the same behaviour of the fidelity amplitude as the one shown for quantum maps, in particular 
we could clearly observe the Lyapunov regime. 
Therefore we have made a step forward towards the settlement of this longstanding problem: we showed an
example of a realistic system where
the Lyapunov regime is observed, independently of the type of initial 
state,  if the appropriate quantity -- the fidelity amplitude -- is considered.

Additionally, we have 
shown that these regimes were also manifested in the behaviour of the IPR, which is a very relevant quantity in the study of localisation 
and quantum chaos at the level of the structure of eigenstates. 
Finally, for completeness we studied  the dynamics of the fidelity amplitude using the semiclassical DR approximation and showed a good agreement with the quantum results. 
This suggests that, since fidelity is also related to the Fourier transform of the work probability distribution after a quench \cite{silva08},  further 
insight into this issue can be obtained by considering tools such as the semiclassical DR approximation.\\
\label{finalremarks}
\acknowledgments
The authors have received funding from CONICET (Grants No. PIP 114-20110100048 
and No. PIP 11220080100728)
ANPCyT (Grants No. PICT-2010-02483, No. PICT-2013-0621, No.PICT 2010-1556) 
and UBACyT.

\end{document}